\documentclass[12pt,preprint]{aastex}
\usepackage{emulateapj5}
\newcommand{\mincir}{\raise
-2.truept\hbox{\rlap{\hbox{$\sim$}}\raise5.truept\hbox{$<$}\ }}
\newcommand{\magcir}{\raise
-2.truept\hbox{\rlap{\hbox{$\sim$}}\raise5.truept\hbox{$>$}\ }}
\newcommand{\minmag}{\raise
-2.truept\hbox{\rlap{\hbox{$<$}}\raise6.truept\hbox{$<$}\ }}
 
\usepackage{graphicx}
\newcommand{\be}{\begin{equation}}
\newcommand{\ee}{\end{equation}}

\def\arcdeg{\hbox{$^\circ$}}

\newenvironment{inlinefigure}{%
\def\@captype{inlinefigure}%
\noindent\begin{minipage}{\linewidth}\begin{center}}
{\end{center}\end{minipage}\smallskip}
\makeatother
\shorttitle{Bias $z$-evolution in the $\Lambda$CDM Cosmology}
\shortauthors{Basilakos, Plionis \& Ragone-Figueroa}

\begin{document}


\title{The Halo Mass-Bias Redshift Evolution in the $\Lambda$CDM Cosmology}
\author{S. Basilakos\altaffilmark{1}, M. Plionis\altaffilmark{2,3}, 
C. Ragone-Figueroa\altaffilmark{4,5}}

\altaffiltext{1}{Research Center for Astronomy \& Applied Mathematics,
Academy of Athens, Soranou Efessiou 4, GR-11527 Athens, Greece}
\altaffiltext{2}{Institute of Astronomy \& Astrophysics, 
National Observatory of Athens, Palaia Penteli 152 36, Athens, Greece}
\altaffiltext{2}{Instituto Nacional de Astrof\'{\i}sica \'Optica y
Electr\'onica, AP 51 y 216, 72000, Puebla, Pue, M\'exico}
\altaffiltext{4}{Grupo IATE-Observatorio Astron\'omico, Laprida 854, C\'ordoba,
Argentina}
\altaffiltext{5}{Consejo de Investigaciones Cient\'{\i}ficas y
T\'ecnicas de la Rep'/ublica Argentina, C\'ordoba, Argentina}

\begin{abstract} 
We derive an analytic model for the redshift evolution of
linear-bias, allowing for interactions and merging of the mass-tracers,
by solving a second order differential equation
based on linear perturbation theory and the Friedmann-Lemaitre 
solutions of the cosmological field equations.
We then study the halo-mass dependence of the bias 
evolution, using the dark matter halo distribution 
in a $\Lambda$CDM simulation in order to calibrate
the free parameters of the model.
Finally, we compare our theoretical predictions
with available observational data and find a good agreement. 
In particular, we find that the bias of optical QSO's evolve differently
 than those selected in X-rays and that their corresponding typical dark
 matter halo mass is $\sim 10^{13} \; h^{-1} \; M_{\odot}$ and
 $\magcir 5 \times 10^{13} \; h^{-1} \; M_{\odot}$, respectively.

\end{abstract}


\section{Introduction} 
The distribution of matter on 
large scales, based on different extragalactic objects, 
can provide important constraints on models of cosmic structure formation. 
However, a serious problem that hampers such a straight forward 
approach is our limited knowledge of how luminous matter traces the
underlying mass distribution. 
It is well known that the
large scale clustering pattern of different extragalactic objects
(galaxies, AGN, clusters, etc) trace the underlying dark matter distribution
in a biased manner (Kaiser 1984; Bardeen et al. 1986). 
Such a biasing is assumed to be statistical 
in nature; with galaxies and clusters being identified as high peaks
of an underlying, initially Gaussian, random density field.
Furthermore, the biasing of galaxies with respect to the dark matter 
distribution was
also found to be a necessary ingredient of Cold Dark Matter (CDM) 
models of galaxy formation in order to reproduce the observed galaxy 
distribution (eg. Davis et al. 1985; Cole \& Kaiser 1989; 
Benson et al. 2000).

Furthermore, the bias redshift evolution
is very important in order to relate observations with models of
structure formation and it has been shown that the 
bias factor, $b(z)$, is a monotonically increasing function of redshift. 
Indeed, Mo \& White (1996) and Matarrese et al. (1997) have developed a 
model for the evolution of the correlation bias, defined as the ratio of the 
halo to the mass correlation function, the so called galaxy merging bias 
model. 
This model takes into account, via the Press-Schechter 
formalism, the collapse of different mass halos at the different epochs.
In this framework and at high $z$'s one finds, in the EdS universe, that
$b(z)-1 \propto (1+z)$  (see also Sheth et al. 2001). 
According to Matarrese et al (1997), 
if the only halos that exist at any epoch are those that have just
formed via a merging process of smaller halos 
then the bias evolution, in the high density peak limit,   
becomes $b(z) \propto (1+z)^{2}$ (see also Bagla 1998).  
There are many studies of the merging bias model in the context of a 
$\Lambda$ cosmology (e.g. Jing 1998; Sheth \& Tormen 1999; 
Sheth, Mo \& Tormen 2001), while the bias has been found to be a key 
ingredient of the halo model (e.g. Peacock \& Smith 2000; Seljak 2000; 
Ma \& Fry 2000).

The so-called galaxy conserving bias model (Nusser \& Davis 1994; 
Fry 1996; Tegmark \& Peebles 1998; Hui \& Parfey 2007) 
assumes only that the different mass-tracer (acting as ``test particles'') 
fluctuation field is proportional to that of the underlying mass 
and it predicts a bias evolution according to:
$b(z)=1+(b_{0}-1)(1+z)$ for $\Omega_{\rm m}=1$, 
where $b_{0}$ is the bias factor at the present time. 
Coles, Melott \& Munshi (1999) have developed a bias model within the 
hierarchical clustering scenario which gives interesting scaling 
relations for the galaxy bias.
Another approach was 
proposed by Basilakos \& Plionis (2001; 2003), in which the linear 
bias evolution was
described via the solution  of a second order differential equation, 
derived using linear perturbation theory and the basic cosmological
equations. 
Also, in this model the mass tracer population is assumed to be 
conserved in time.

In this paper we extend the original Basilakos \& Plionis linear and 
scale-independent bias
evolution model to include the effects of an evolving mass-tracer 
population, possibly due to 
interactions and merging. We then derive the dependence of the 
linear-bias evolution on halo mass, with the help of 
N-body simulations,
in the framework of the concordance $\Lambda$CDM cosmological model. 
Additionally we compare our model with observations.

\section{Linear Bias Evolution Model}
In this section we describe our general linear bias evolution model 
which is based on linear perturbation theory in the matter 
dominated epoch (eg. Peebles 1993), either assuming that the
mass-tracer population is conserved in time or that it evolves
according to a $(1+z)^{\nu}$ law.

\subsection{Case 1: Conservation of Mass-tracer Population}
In Basilakos \& Plionis (2001; 2003) we have assumed 
that in the evolution of the linear bias 
the effects of non-linear gravity and 
hydrodynamics (merging, feedback mechanisms etc) can be ignored 
(eg. Fry 1996; Tegmark \& Peebles 1998; 
Catelan et al. 1998). Then, using the linear perturbation theory 
we obtained a second order differential equation which describes the 
evolution of the linear bias factor, $b$, between the background
matter and the mass-tracer fluctuation field:
\be\label{eq:hdif} 
\ddot{y}\delta + 2(\dot{\delta} + H \delta) \dot{y} + 4 \pi G
\rho_{\rm m} \delta y =0 \;,
\ee
where $b=y+1$ and $\delta(t) \propto D(t)$. 
Note that $D(t)$ is the linear growing mode
(scaled to unity at the present time), useful expressions 
of which can be found for the $\Lambda$ cosmology in
Peebles et al. (1993):
\be\label{eq:24}
D(z)=\frac{5\Omega_{\rm m} E(z)}{2}\int^{\infty}_{z} \frac{(1+x)}{E^{3}(x)} 
{\rm d}x \;,
\ee
and for the quintessence
models in Silveira \& Waga (1994), Wang \& Steinhardt (1998) and
Basilakos (2003).
The solution of eq.(\ref{eq:hdif}) 
provided our bias evolution model (Basilakos \& Plionis 2001). The
solution for the different cosmological model enter through the
different behavior of $D(t)$ and $H(t)$.

In order to transform equation (\ref{eq:hdif}) 
from time to redshift, we utilized the following expressions:
\be\label{eq:5}
\frac{dt}{dz}=-\frac{1}{H_{0}E(z)(1+z)} 
\ee
\be
E(z)=\left[ \Omega_{\rm m}(1+z)^{3}+\Omega_{\Lambda}\right]^{1/2} \;\;,
\ee
with $\Omega_{\rm m}$ being the 
density parameter at the present time, which satisfies
$\Omega_{\Lambda}=1-\Omega_{\rm m}$.
Taking into account the latter transformations, the 
basic differential equation for the evolution of the linear
bias parameter takes the following form:
\be\label{eq:gen2}
\frac{{\rm d}^{2} y}{{\rm d} z^{2}}-P(z)\frac{{\rm d} y}{{\rm d} z}+
Q(z)y=0 \;\; 
\ee
with relevant factors, 
\be\label{eq:ff1}
P(z)=\frac{1}{1+z} - \frac{1}{E(z)}\frac{{\rm d}E(z)}{{\rm d}z}
-\frac{2}{D(z)}\frac{{\rm d}D(z)}{{\rm d}z}
\ee
and
\be\label{eq:ff2}
Q(z)=\frac{3\Omega_{\rm m} (1+z)}{2E^{2}(z)} \;\; .
\ee

The general solution as a function of redshift 
for flat cosmological models ($\Omega_{\rm m}+\Omega_{\Lambda}=1$) 
was found to be (Basilakos \& Plionis 2001):    
\be\label{eq:final} 
b(z)= {\cal C}_{1} E(z)+{\cal C}_{2} E(z)I(z)+1 \;\;, 
\ee
with
\be\label{eq:88} 
I(z)=\int_{z}^{\infty} \frac{(1+x)^{3}}{E^{3}(x)}  {\rm d}x\;\;.
\ee
The integral of equation (\ref{eq:88}) is elliptic and therefore
its solution, in the redshift range $[z,+\infty)$, can be expressed 
as a hyper-geometric function:
\be
I(z) \propto (1+z)^{-1/2} 
F\left[\frac{1}{6},\frac{3}{2},\frac{7}{6},-\frac{\Omega_{\Lambda}}
{\Omega_{\rm m} (1+z)^{3}} \right] \;\; . 
\ee
Note that the first term in the right-hand side of eq.(\ref{eq:final}), 
which is the dominant one, has an approximate 
redshift dependence $\sim (1+z)^{3/2}$ while the second term has 
a dependence $\sim (1+z)$.

In Basilakos \& Plionis (2001) we compared this bias evolution 
model with the halo merging model (eg. Mo
\& White 1996; Matarrese et al. 1997) for $z\le 3$ 
and found a very good consistency, once we fitted the integration 
constants ${\cal C}_{1},{\cal C}_{2}$ by evaluating our model to two
different epochs. Also in Basilakos \& Plionis (2003) we compared 
our $\Lambda$CDM solution,
evaluated at $z=0$ and $z=3$ using the Hubble Deep Field (HDF) 
results (Arnouts et al. 2002), with the 
Matarrese et al (1997) model and again found a good consistency. 

However, the consistent comparison of our bias evolution model 
with the available observational and numerical data is only 
after we normalize our model to two different epochs, given by the
data. In other words there is no independent 
normalization of our bias evolution curves.

\subsection{Case 2: Evolving Mass-tracer Population}
The assumption used in the previous section, that  
the mass-tracer number density is conserved in time, should be a gross 
oversimplification, since in most galaxy formation models 
it is expected that interactions and merging is 
very important at early epochs.

We now drop this assumption by allowing 
a contribution from the corresponding 
interactions among the mass tracers and thus an evolution of the
mass-tracer population.
We derive again the corresponding equation (1), starting from the continuity
equation and introducing an additional time-dependent term, $\Psi(t)$, which we
associate with the effects of interactions and merging of the mass tracers.
We also make the same assumption, as in our original formulation,
that the tracers and the underlying mass distribution
share the same velocity field. Then:
\be
\dot{\delta} + \nabla u \simeq 0 \;\; \mbox{\rm and} \;\; \dot{\delta}_g +
\nabla u +\Psi(t) \simeq 0 \;,
\ee 
from which we obtain 
\be 
\dot{\delta} - \dot{\delta}_g=\Psi \;.
\ee 
Although we do not have a fundamental theory to model the 
time-dependent $\Psi(t)$ function, 
according to the notations of Simon (2005), it appears to depend 
on the tracer number 
density and its logarithmic derivative as well as on the tracer overdensity:
$\Psi(t) \propto \Psi(\bar{n}, (1+\delta_g) d \ln {\bar n}/dt)$ (see eq. 10 of 
Simon 2005 and our appendix).

Now, since we are dealing with 
linear biasing, we have $\delta_g=b \delta$ and using $b=y+1$, we get
that $d(y \delta)/dt=-\Psi$. Differentiating twice we then have:  
$\ddot{y} \delta + 2 \dot{y} \dot{\delta} + y \ddot{\delta} =-\dot{\Psi}$.
Solving for $y \ddot{\delta}$, using the fact that $y
\dot{\delta} = -\dot{y}\delta-\Psi$ and utilizing the differential
time-evolution equation of $\delta$ (eg. Peebles 1993):
\be
\ddot{\delta}+ 2H\dot{\delta}-4 \pi G\rho_{\rm m} \delta =0
\ee
we finally obtain:
\be
\ddot{y}\delta + 2(\dot{\delta} + H \delta) \dot{y} + 4 \pi G
\rho_{\rm m} \delta y =-2H\Psi-{\dot \Psi}
\ee
which is the corresponding equation (1) for the case of interactions
among the tracers.

If we transform the latter equation from time to redshift
the basic differential equation becomes:
\be\label{eq:gen2}
\frac{{\rm d}^{2} y}{{\rm d} z^{2}}-P(z)\frac{{\rm d} y}{{\rm d} z}+
Q(z)y=f(z) 
\ee
with
\be
f(z)=\frac{(1+z)\Psi^{'}(z)E(z)-2E(z)\Psi(z)}{H_{0}D(z)} 
\ee
where the prime denotes derivative with respect to redshift.

Therefore, if one is able to find a partial solution $y_{p}$ of 
eq. (\ref{eq:gen2}), 
then due to the fact that we know already the solutions [$E(z)$ and $E(z)I(z)$ 
in section 2.1] for the homogeneous case, the general solution for the
interactive case should satisfy the following formula:
\be
y(z)= {\cal C}_{1} E(z)+{\cal C}_{2} E(z)I(z)+y_{p}(z) \;\;.
\ee\label{eq:final1}
Using some basic elements of the differential equation 
theory, the corresponding partial solution becomes
\begin{eqnarray}
y_{p}(z)=
E(z) \int_{0}^{z}\frac{f(x)E^{2}(x)I(x)}{(1+x)^{3}}{\rm d}x- \nonumber \\ 
E(z)I(z)\int_{0}^{z} \frac{f(x)E^{2}(x)}{(1+x)^{3}} {\rm d}x \;\;.
\end{eqnarray}
It is obvious that if the interaction among the 
tracers is negligible ($\Psi\simeq 0$) then  
$y_{p}\simeq 0$, as it should.

In order to proceed we need to define the functional form 
of the interaction term $\Psi(z)$ which is not an 
easy task to do.
Due to the absence of a physically well-motivated fundamental theory, 
we parametrize the interaction term utilizing a standard evolutionary form:
\be
\Psi(z)=AH_{0}(1+z)^{\nu} 
\ee
implying that
\be
f(z)=A(\nu-2)\frac{(1+z)^{\nu}E(z)}{D(z)}\;,
\ee
where $A$ and $\nu$ are positive parameters (to be determined
from simulations see next section). This is a reasonable approach since it can be
shown that indeed our $\Psi(z)$ term has such a dependence (see appendix).
From a mathematical point of view,
we have that for $\nu>2$ the bias evolution becomes stronger than 
in the case of no interactions, especially at high redshifts,
which means that due to the merging processes the halos (of some particular 
mass) correspond to higher peaks of the underlying density field with respect 
to equal mass halos in the non-interacting case. 
On the other hand, the $\nu< 2$ case corresponds to the destruction of halos
of a particular mass, which results into a lower-rate of bias evolution with 
respect to the non-interacting case.
Now, for the limiting case with $\nu=2$ we obtain $y_p=0$, 
implying no contribution of the interacting term to the bias evolution 
solution, as in the case with $\Psi(t)=0$, which can be interpreted as 
the case where the destruction and creation processes are counter-balanced. 

\section{Parametrizing the Bias Evolution Model using N-body 
Simulations}
Our analytical approach gives a family of bias
curves with four unknown parameters (${\cal C}_{1},{\cal C}_{2},A,\nu$).
In order to obtain the behavior of $b(z)$ we need to somehow 
evaluate these constants for different halo-masses
(since different halo-masses result in different $b(0)$ and different 
rate of bias evolution) and we do so by using a $\Lambda$CDM, dark matter
only, simulation.

\subsection{Simulation Data}
The simulation volume is of a 500 $h^{-1}$ Mpc cube in which a random
realization of the concordance $\Lambda$CDM
$(\Omega_{\rm m}=0.3$, $\Omega_{\Lambda} = 0.7$, $h=0.72$ and 
$\sigma_8=0.9$) power spectrum was generated with $512^3$ particles
(Ragone-Figueroa \& Plionis 2007).
The simulation was performed with an updated version of the parallel
Tree-SPH code GADGET2 (Springel 2005). 
The particle mass is $m_{\rm p}\ge 7.7\times
10^{10}\,h^{-1}\,M_{\odot}$ comparable to the mass of one single
galaxy. 
The halos are defined using a FoF algorithm with a linking
length $l=0.17\langle n\rangle^{-1/3}$, where $\langle n \rangle$ 
is the mean particle density. This linking length corresponds to an 
overdensity $\simeq 330$ at the present epoch ($z=0$). 

We estimate the bias redshift evolution of the different dark mass halos, 
with respect to the underlying matter distribution,
by measuring their relative fluctuations in spheres of radius 
8 $h^{-1}$ Mpc, according to: 
\be
b(M,z)=\frac{\sigma_{8, h}(M,z)}{\sigma_{8, m}(z)} \;,
\ee 
where the subscripts $h$ and $m$ denote halos and 
the underlying mass, respectively.
The values of $\sigma_{8,h}(M,z)$, for halos of mass $M$, are computed
at different redshifts, $z$, by:
\be
\sigma^2_{8,h}(M,z)=\left\langle \left(\frac{N-\bar{N}}{\bar{N}} 
\right)^2\right\rangle-\frac{1}{\bar{N}} \;,
\ee 
where $\bar{N}$ is the mean number of such halos in spheres of 8
$h^{-1}$ Mpc radius and the factor $1/\bar{N}$ is the expected
Poissonian contribution to the value of $\sigma^2_{8,h}$.
Similarly, we estimate at each redshift the value of the underling
mass $\sigma_{8,m}$.
In order to measure $\sigma^2_{8,j}$ we randomly place 
$N_{\rm rand}$ sphere centers in the simulation volume, such that the sum of 
their volumes is equal to $1/8$ of the simulation volume ($N_{\rm rand}\simeq 
7500$).  This is to ensure that we are not oversampling the available 
volume, in which case we would have been multiply sampling the same
halo or mass fluctuations. The uncertainties reflect 
the dispersion in the distribution of
$\sigma^2_{8,j}$, taking 20 bootstrap resamplings of the halo sample. 
Note, that these errors do not take into account the uncertainties
introduced by cosmic variance, although for the majority of 
the halo mass ranges investigated it should be negligible, 
due to the size of the simulation volume. Nevertheless, we verified this by 
investigating the effects of cosmic variance on the value of $\sigma_8$ 
for volumes 8 times smaller than our simulation box and found them to be 
significantly smaller than the corresponding bootstrap errors.

For our present analysis we use 10 different redshift 
snapshots, spanning the range: $0\le z \le 5$ and
five different halo mass intervals: 
\begin{itemize}
\item Sample $S_{1}$: $7.7\times 10^{11}h^{-1}M_{\odot} \le M \le
10^{13}h^{-1}M_{\odot}$ 
\item Sample $S_{2}$: $1\times 10^{13}h^{-1}M_{\odot}<M \le 3\times
10^{13}h^{-1}M_{\odot}$ 
\item Sample $S_{3}$: $3\times 10^{13}h^{-1}M_{\odot} < M \le 6\times
10^{13}h^{-1}M_{\odot}$ 
\item Sample $S_{4}$: $6\times 10^{13}h^{-1}M_{\odot} < M \le 1\times
10^{14}h^{-1}M_{\odot}$ 
\item Sample $S_{5}$: $1\times 10^{14}h^{-1}M_{\odot} < M \le 4\times
10^{14}h^{-1}M_{\odot}$ 
\end{itemize}

In Figure 1 we present with dots the simulation $b(M,z)$ results
for the different subsamples
(solid points $S_{1}$, open points $S_{2}$, crosses $S_{3}$, 
solid squares $S_{4}$ and 
open squares $S_{5}$). As it is expected, the biasing 
is a monotonically increasing 
function of redshift with its evolution being significantly 
stronger for larger halo masses (eg. Kauffmann et al. 1999; 
Somerville et al 2001 and references therein). By solid curves we
present our bias evolution models (case's 1 and 2 at the left and
right panels, respectively), which have been fitted to the N-body
results (see below).

\begin{inlinefigure}
\plotone{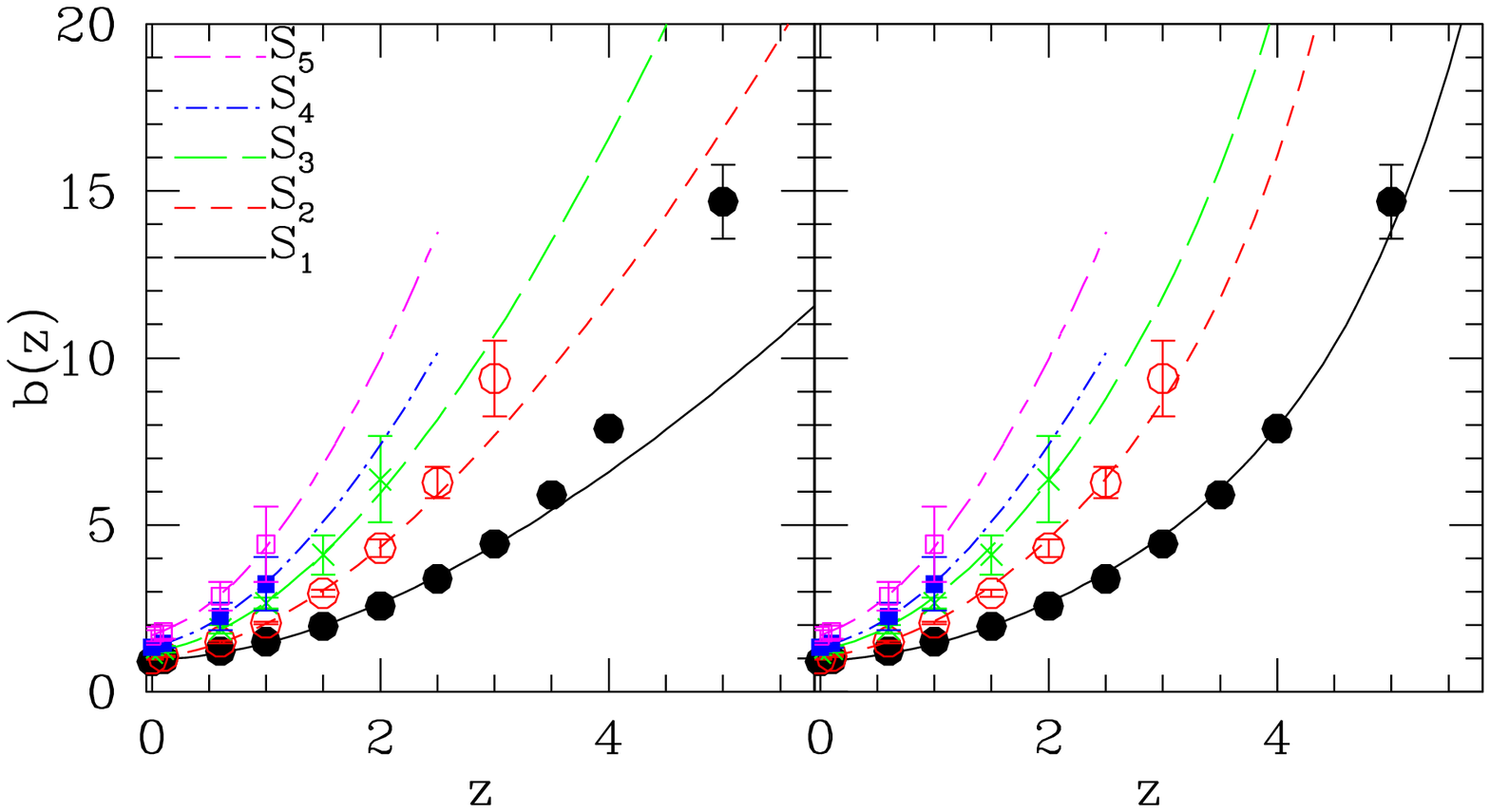}
\figcaption{ The bias $z$-evolution for different halo masses in a
concordance $\Lambda$CDM simulation (points). Continuous lines
represent our bias evolution model, fitted to the simulation results:
({\em Case 1} is shown in the left panel and {\em Case 2} in the
right panel).
Note that the different continuous lines corresponds to different halo
masses and thus to different values of the fitted model parameters,
shown in Table 1.}
\end{inlinefigure}

\subsection{Fitting our Model $b(z)$}
Firstly we fit to the simulation data our 
({\em Case 1}) model, where the mass tracers population is conserved 
in time.
To do so we utilize a standard $\chi^{2}$ 
likelihood procedure and compare the measured bias evolution, $b(M,z)$,
with that expected from our theoretical approach (see eq. \ref{eq:final}).
We define the likelihood estimator\footnote{Likelihoods
are normalized to their maximum values.} as:
${\cal L}({\bf c}) 
\propto {\rm exp}[-\chi^{2}({\bf c})/2]$
with:
\be
\chi^{2}({\bf c})=
\sum_{i=1}^{n} \left[ \frac{b^{i}(M,z)-b^{i}({\bf c},z)}
{\sigma^{i}}\right]^{2} \;\;,
\ee 
where ${\bf c}$ is a vector containing the  
parameters that we want to fit and $\sigma_{i}$ 
is the derived uncertainty (see previous section).
Note, that in this case ${\bf c}=({\cal C}_{1},{\cal C}_{2})$.
We sample the various parameters as follows:
${\cal C}_{1} \in [0,25]$ and 
${\cal C}_{2} \in [-2,4]$ 
in steps of 0.01. 

In Fig. 1 (left panel), we plot our derived 
bias evolution fits (lines) to the simulation data (points) 
to find that indeed they compare very well for $z\mincir 2.5 - 3$.
However, it appears that at larger redshifts the
simulation halo bias evolves more rapidly than what our 
({\em Case 1})
model predicts, especially for the low mass halos 
(see $S_{1}$, $S_{2}$ and $S_{3}$ subsamples), which
of course should be attributed to the assumption of halo
number density conservation. 
However, it is evident that this model describes well
the bias evolution for the high mass halos (see samples 
$S_{4}$ and $S_{5}$), since such high-mass halos form quite late in the 
evolution of structure formation processes.

These results strongly indicate that the interactions among the 
relative low mass tracers at large redshifts play an important role in 
the evolution of halo biasing. 
We now perform our likelihood analysis for the {\em Case 2} model, ie.,
using the corresponding vector:
${\bf c}=({\cal C}_{1},{\cal C}_{2},A,\nu)$.
We sample the $(A,\nu)$ parameters as follows:
$A \in [10^{-3},10^{-2}]$ in steps of $10^{-3}$; and 
$\nu \in [2,3]$ in steps of 0.02. The procedure is applied only
for the lower mass halos (see $S_{1}$, $S_{2}$ and $S_{3}$ subsamples).
The resulting best fit parameters
and the zero-point bias, $b_{0}$
\footnote{For $z=0$ we get $b_{0}=b(0)=1+{\cal C}_{1}+{\cal C}_{2}I(0)$.
Note that for the concordance model
$\Omega_{\rm m}=1-\Omega_{\Lambda}=0.3$ 
we have $I(0)\simeq 9.567$}, for the five
subsamples, are presented in Table 1, while we plot 
the results, as continuous lines, in the right panel of figure 1. 

Our {\em Case 2} model fits extremely well the numerical simulation
results even at high redshifts, which implies that our modification of 
the test-particle bias solution, to take into account interactions
among mass-tracers,  is an extremely good approximation to
the more elaborate halo bias evolution solutions, 
based on the Press-Schechter formalism. 
Note also that the derived values of the 
constants of integration, $({\cal C}_{1},{\cal C}_{2})$, 
are very similar in both
of our models ({\em Case 1} and {\em Case 2}).

\begin{inlinefigure}
\epsscale{1.}
\plotone{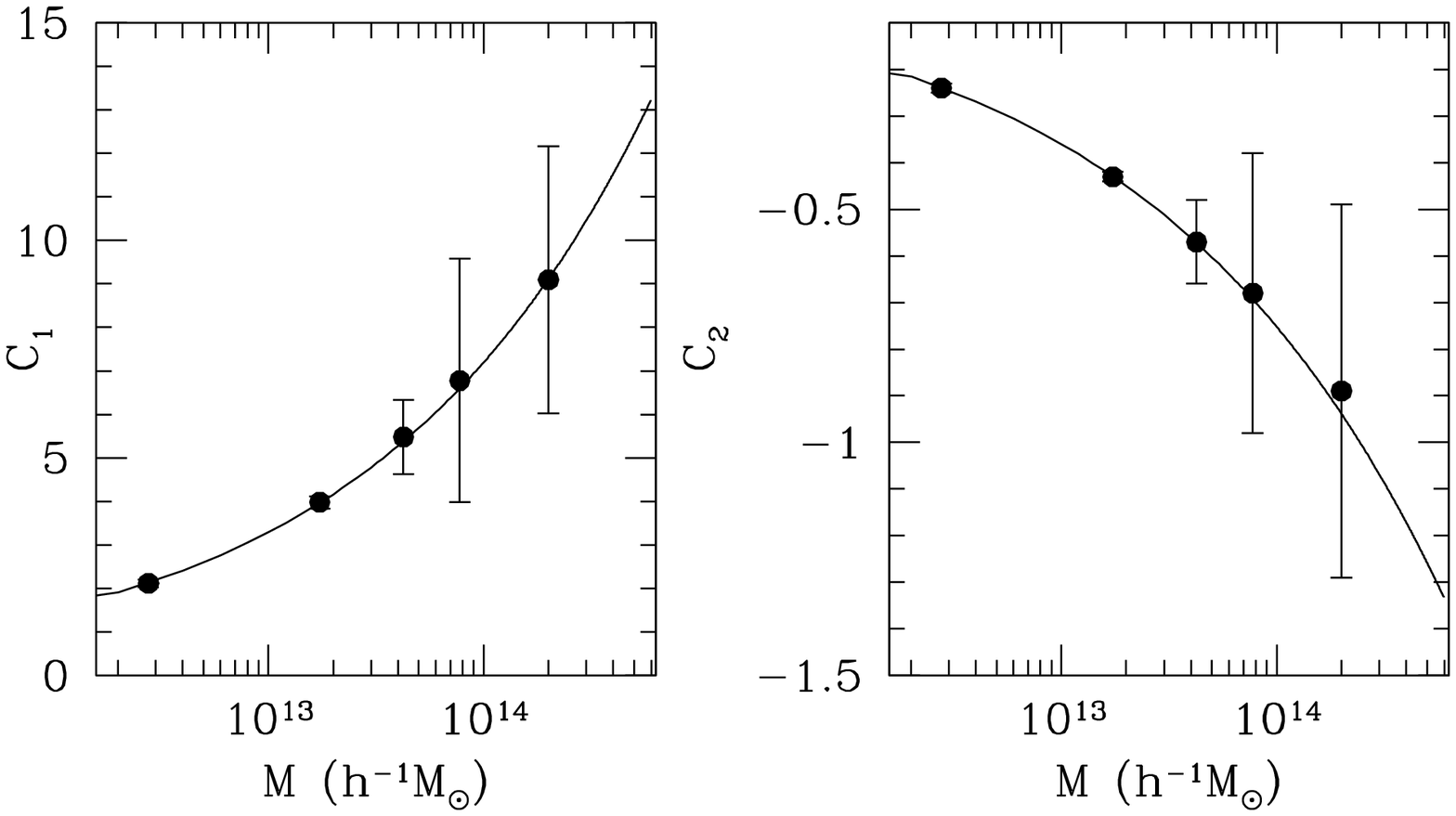}
\figcaption{The relation between 
the measured constants and the median of each halo mass interval
(${\cal C}_{1}$ and ${\cal C}_{2}$ in the {\em right} and {\em left}
panels, respectively).
The corresponding fits, in both panels, are given by eq.(24).}
\end{inlinefigure}

We would like to caution the reader that the fitted parameters, tabulated in 
Table 1, are based on a concordance $\Lambda$CDM simulation with a high
$\sigma_8 \; (=0.9)$ normalization. We have investigated the sensitivity of 
these parameters to the power-spectrum normalization by running a {\em WMAP3} 
normalized ($\sigma_8=0.75$) simulation (but of a smaller box size: 
$L=250 \; h^{-1}$ Mpc) and found small differences and only for the 
${\cal C}_1$ and $b(0)$ parameters. We find for the {\em WMAP3} 
normalization case that the value of ${\cal C}_1$ and $b(0)$ are higher by 
$\sim 4\%$ and $\sim 6\%$, respectively, with respect to those presented in 
Table 1. On the other hand, the choice of the value of $\Omega_{\rm m}$
affects only the constant ${\cal C}_2$, which for the case of different
flat cosmologies should be multiplied by a factor $(\Omega_{\rm m}/0.3)^{3/2}$ 
(see Basilakos \& Plionis 2001).

Now, in order to derive a generic halo-mass dependent formula of our bias
evolution model, within the $\Lambda$CDM model,
we correlate the constants ${\cal C}_{1}, {\cal C}_{2}$, 
with the median mass of the corresponding halo sub-samples
and find the following fits (shown in figure 2):

\be
{\cal C}_{1,2}(M)\simeq \alpha_{1,2} \left( \frac{M}{10^{13}h^{-1}M_{\odot}}
\right)^{\beta_{1,2}} \;,
\ee\label{eq:fitC}
where $\alpha_{1}=3.29\pm 0.21$, $\beta_{1}=0.34\pm 0.07$
and $\alpha_{2}=-0.36\pm 0.01$, $\beta_{2}=0.32\pm 0.06$
are the corresponding best values with their 
$1\sigma$ uncertainties. 

Finally, we need to emphasize that with our current model,
assuming also that the extragalactic mass tracers (clusters, galaxies, 
X-ray sources, etc.) are hosted by a halo of a given mass, 
it is straightforward to derive analytically their bias evolution behavior.

\begin{inlinefigure}
\epsscale{1.04}
\plotone{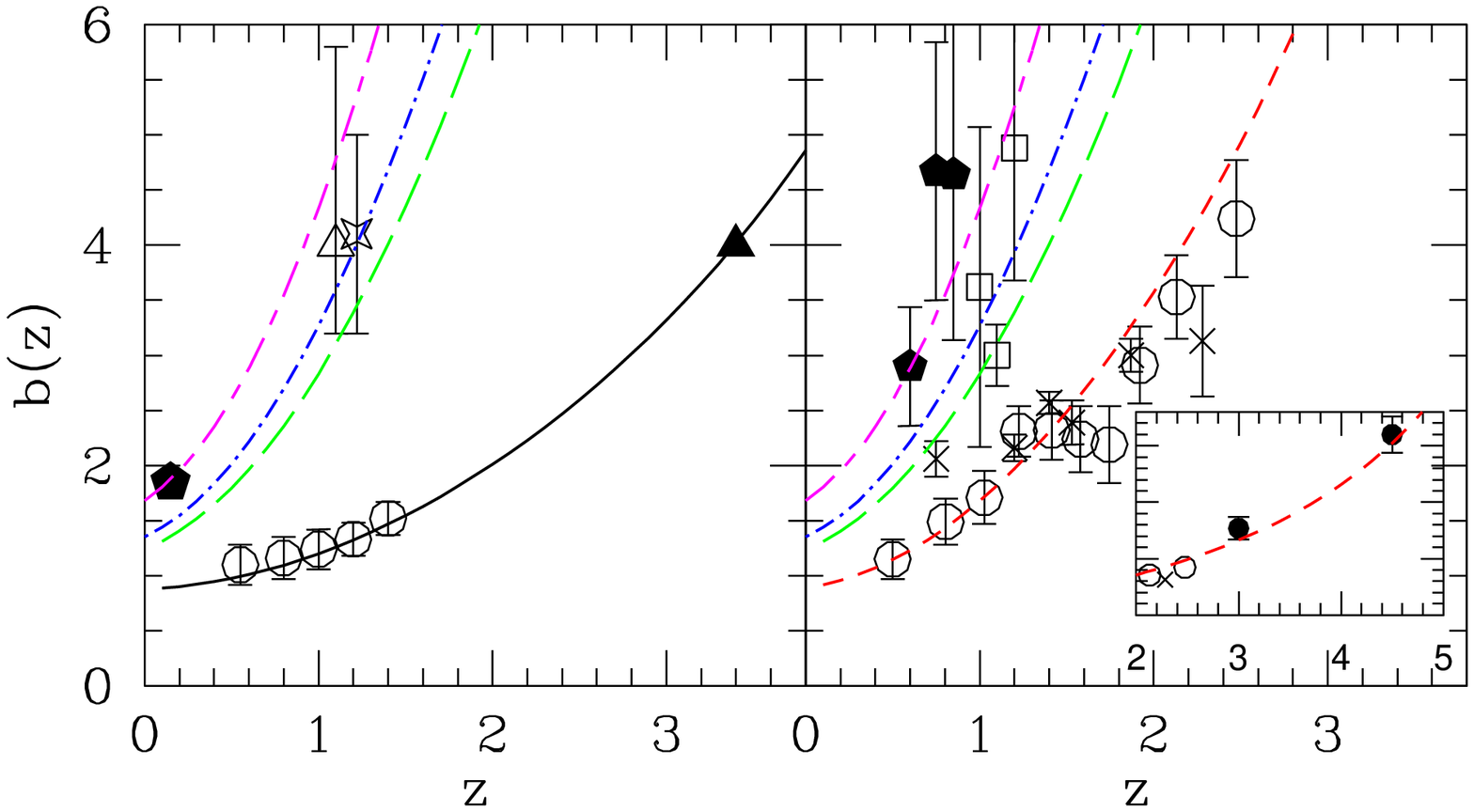}
\figcaption{
Comparison of our $b(z)$ model with different observational data.
Different line types represent different 
halo masses (see section 3.1). 
{\em Left Panel:} optical galaxies (open points), 
Lyman break galaxies (solid triangle), EROs (star), DRGs (open triangle)
and 2dF radio sources (filled pentagon).
{\em Right Panel:} optically selected quasars (open points and crosses), 
soft and hard X-ray point sources (open squares and solid
diamonds). In the insert we plot, as solid points, the high-$z$ SSRS 
DR5 QSOs and the same $b(z)$ 
model that fits their lower redshift counterparts (ie., $M \simeq 
10^{13} \;h^{-1} \; M_{\odot}$).
}
\end{inlinefigure}

\section{Comparison with Observations \& other Bias Models}
We investigate the extent to which our bias-evolution model can fit recent
observational data of different extragalactic sources and which is the  
DM halo mass for which such a comparison is successful. 
In performing such a comparison we implicitly assume that, on average, 
each class of extragalactic sources is related to DM haloes of some, specific,
mass.
Therefore, to the extent to which our model fits the data, we will be
able to identify the (average) mass of the DM halo which the
different extragalactic sources inhabit.
We will also compare our model with other recent bias evolution models to show
the extent to which they compare.

In the left panel of figure 3 we compare our theoretical prediction
with observations using optical galaxies (open circles) from 
the VIMOS VLT Deep Survey (Marinoni et al. 2005). It is quite 
evident that our model (solid line), corresponding to a DM
halo mass of $\sim 10^{12} \;h^{-1} M_{\odot}$, 
represents extremely well the $z$-dependence of the optical
galaxy bias. It is quite interesting that the same $b(z)$
curve fits accurately the Lyman-break galaxies (solid triangle) 
at a redshift $z\ge 3$ (Steidel et al. 1998;
Adelberger et al. 1998; Kashikawa et al. 2006). 

Roche et al. (2002) using the ELAIS N2 field  
and Georgakakis et al. (2005) based on the PHOENIX deep survey
find that the correlation length of the extremely red galaxies (EROs)
lies in the interval $r_{0}\simeq 10-17 h^{-1}$ Mpc for a
slope $\gamma=1.8$. Thus, the corresponding EROs bias value, derived using:
$$
b(z)=\left(\frac{ r_{0} }{r_{0,m}} \right)^{\gamma/2} 
D^{3+\epsilon}(z)\;\; \mbox{with} \; \gamma=1.8 \; \& \; \epsilon=-1.2 \;,
$$
is $b\simeq 4.1\pm 0.9$ at a redshift $z\simeq 1.2$ (shown as a star
in the left panel of Fig. 3).  
While, Foucaud et al. (2007) find for the distant red galaxies
(DRGs) in the UKIDSS Ultra Deep Survey, a bias parameter
of $b=4^{+1.4}_{-0.8}$ at a mean redshift of $z\simeq 1$ 
(see open triangle). Our model bias-evolution curve that fits quite well
both of these results (EROs and DRGs) is the one corresponding to a
DM halo mass of
$\sim 7.7\times 10^{13} h^{-1} M_{\odot}$ (dot dashed line). For
comparison we also show $b(z)$ curves for two more halo masses:
$\sim 4.2\times 10^{13} h^{-1} M_{\odot}$ (long dashed line) and
$\sim 2\times 10^{14} h^{-1} M_{\odot}$ (sort long line).

Finally, Maglioccheti et al. (2004) find $b\simeq 1.85$ (see filled pentagon)
for the 2dF radio galaxies at $z\le 0.15$, corresponding to the
model bias of a $\sim 10^{14} h^{-1} M_{\odot}$ DM halo.


In the right panel of figure 3 we compare the results of optical and
X-ray selected AGNs with our $b(z)$ models. The bias evolution of
optical quasars by Croom et al. (2005) and Myers et al. (2007) 
based on the 2dF QSOs (open circles)
and the SDSS DR4 (crosses) survey, respectively, are well approximated by
our $b(z)$ model for a DM halo of 
$\sim 10^{13}h^{-1}M_{\odot}$ (short dashed line). 
It is interesting to mention that also Croom et al. (2005), 
utilizing the Sheth, Mo \& Tormen (2001) theoretical prescription, found that 
the median DM halo mass,  
for the expected quasar bias evolution, is $\sim 10^{13} h^{-1}M_{\odot}$
(Porciani, Magliocchetti, Norberg 2004;
see also figure 7 of Negrello, Magliocchetti \& de Zotti 2006; 
Hopkins et al. 2007).

\begin{inlinefigure}
\epsscale{1.04}
\plotone{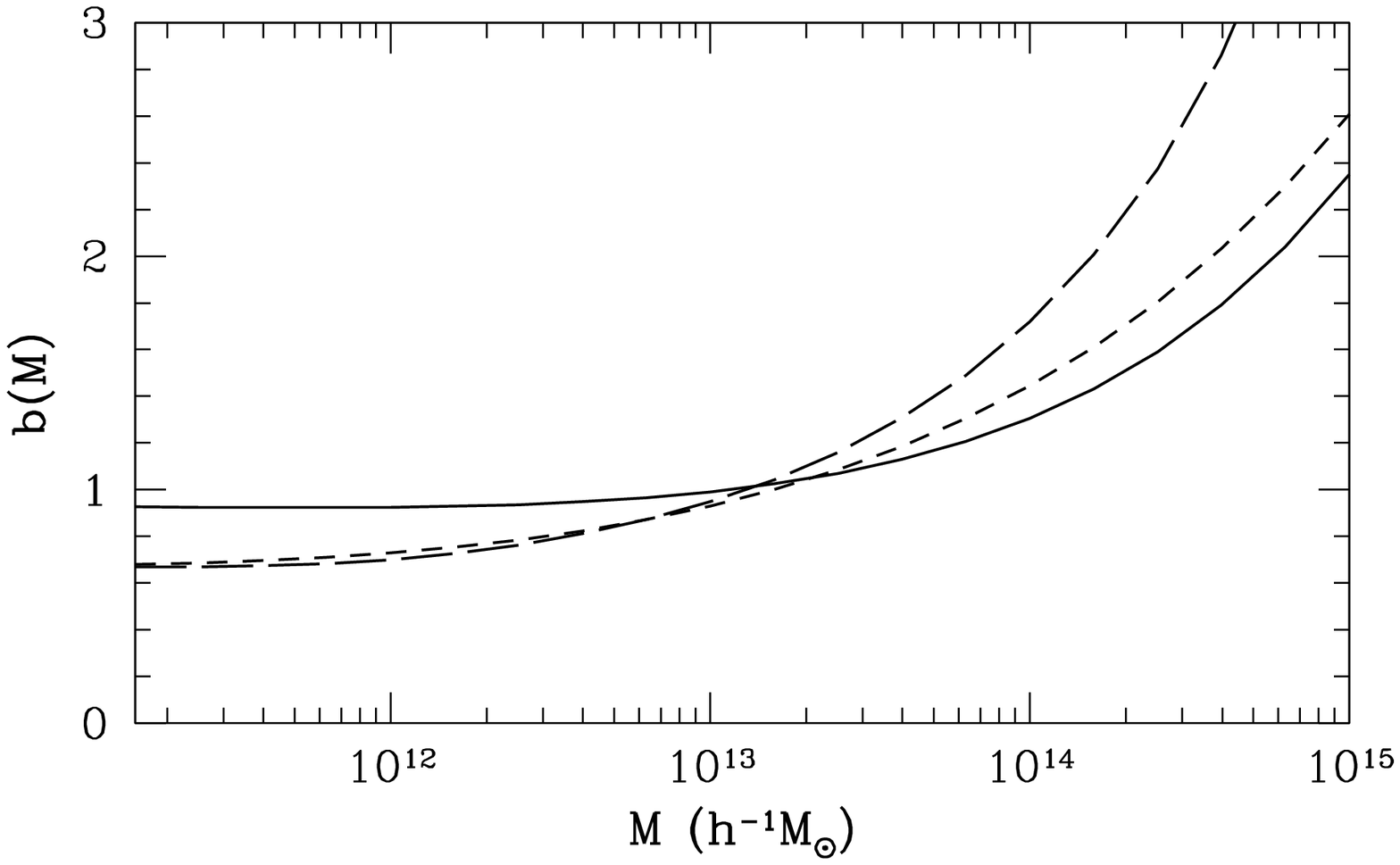}
\figcaption{
The bias parameter as a function of the halo mass at the present 
epoch. The solid line corresponds to our predictions, while  
the dashed and the long dashed lines to those of the Jing (1998) and 
Seljak \& Warren (2004) models, respectively.}
\end{inlinefigure}

Recently, Shen et al. (2007) using the 
SDSS DR5 data found that the high redshift optical quasars 
($2.9\le z \le 5.4$) are more clustered 
than their $z\sim 1.5$ counterparts. The corresponding bias values 
(solid points in the insert plot) are fitted extremely well by the
same model $b(z)$ function as their lower redshift counterparts (ie.,
for a DM halo mass of $\sim 10^{13}h^{-1}M_{\odot}$.

Finally, the relatively strong clustering results of X-ray selected
AGNs, based on a
variety of XMM and Chandra surveys (eg. Gilli et al. 2005; 
Basilakos et al. 2004; 2005; Puccetti et al. 2006; Miyaji et
al. 2007), correspond to bias results shown in the right panel of 
figure 3 as open squares (soft band) and solid diamonds (hard band).
The model $b(z)$ curves that fit these results correspond to halo
masses $M > 5 \times 10^{13} \; h^{-1} \; M_{\odot}$, strongly
suggesting that X-ray and optically selected AGNs do not inhabit the
same DM halos.

From the previous comparisons it is evident that our bias evolution
model ({\em Case 2}) describes well the bias behavior of different 
extragalactic sources.


We now turn to compare 
our generalized test-particle bias model ({\em Case 2}) with some
recent merging bias models. As an example, in figure 4 we 
compare our solution to that of Jing (1998) and Seljak \& Warren (2004)
at $z=0$.
The corresponding functional $b(M)$ forms appear to be quite similar, 
although the predictions of Jing (1998) and Seljak \& Warren (2004), with 
respect to our model, are lower by $\sim 20\%$ and 
higher by $\sim 15\%$ in the low and high-halo mass range, respectively.
It is worth pointing out that for DM halo masses close to those related 
to optical galaxies, the Jing (1998) as well as the 
Seljak \& Warren (2004) models predict a strongly anti-bias picture 
at the present epoch ($b_{0}\sim 0.68$), which appears to be in disagreement
with observational results which indicate that
$b_{0}\sim 1$ (eg. Verde et al. 2002; Lahav et al. 2002).

\begin{inlinefigure}
\epsscale{1.04}
\plotone{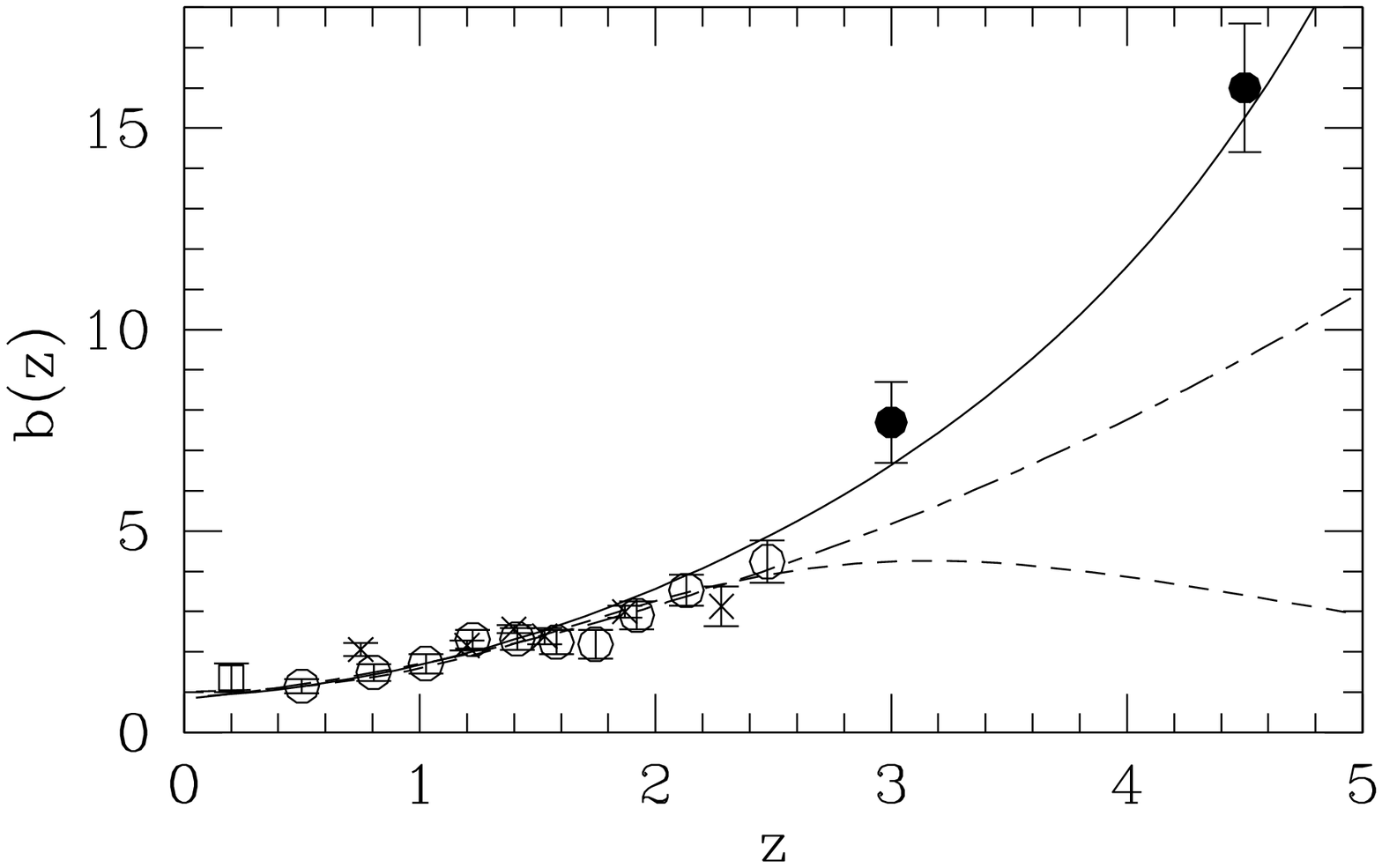}
\figcaption{Comparison of our bias model for a DM halo mass of 
$10^{13} \; h^{-1} \; M_{\odot}$ (solid line) 
with the corresponding Croom et al (2005) empirical form 
(dot-dashed line) and the Hopking et al. (2007) model (dashed line).
The points represent the optical QSO bias
from the 2dF (open points), the SDSS DR4  (crosses) and
the SDSS DR5 (solid points) surveys, while at $z\sim 0.2$ we plot 
the Asiago-ESO/RASS QSO survey results (Grazian et al. 2004).}
\end{inlinefigure}
  
Finally, in figure 5 we again plot the optical QSO bias, as a function of 
redshift (see also Fig. 3), together 
with some bias-evolution model predictions.
Our generalized solution for a halo mass of $M\sim 10^{13}h^{-1}M_{\odot}$
[solid line] can be compared with the empirical formula [dot dashed line] 
derived by Croom et al. (2005; see their equation 15) 
and with the model of Hopkings et al. (2007; see their equation 8) 
[dashed line]. 
The comparison shows that the three curves fit well 
the recent QSO bias evolution ($z<2.5$). At higher redshifts
($z\ge 2.5$) our model evolves significantly 
more than the other two models, and it appears to be the 
only model that fits the recent, observationally derived, high-$z$ 
QSO bias.

\section{Conclusions}
In this work we extend our original Basilakos \& Plionis (2001)
bias evolution model, based on  linear perturbation theory and the 
Friedmann-Lemaitre solutions of the cosmological field equations, 
to take into account the redshift evolution of an interaction term that
could affect the mass-tracer number density. 
Parameterizing our analytical model with 
N-body simulations of the concordance $\Lambda$CDM cosmological model 
($\Omega_{\rm m}=1-\Omega_{\Lambda}=0.3$), we investigate 
the halo-mass dependence of our linear bias evolution model, with or without
the interaction term.

Assuming that the extragalactic mass tracers are hosted 
by a dark matter halo of given mass, we can predict analytically 
their present bias parameter and its $z$-evolution. 
Inversely, fitting our bias evolution model to the observed bias 
of different extragalactic mass-tracers, we can identify the typical 
mass of the DM halo within which they live.

Comparing our theoretical predictions with the observed bias 
of different mass-tracers (galaxies, AGNs, DRGs, EROs, etc)
we find a very good agreement with our bias evolution 
functional form and therefore we also identify the typical 
mass of the DM halo which they inhabit. This has allowed us to 
infer that X-ray and optically selected AGNs inhabit DM halos 
of different mass (with X-ray AGNs being associated with higher 
DM halo masses).

\appendix
\section{Redshift dependence of $\Psi$}
With the aid of the Simon (2005) nomenclature we derive the expected 
redshift dependence of our interaction term $\Psi(z)$. In Simon (2005) 
the related ``sink'' term, $\phi(z)$,  
is modeled as the variation of the tracer number density, $n_{tr}$, given 
by their mass conservation equation:
\be
\frac{dn_{tr}}{dt} + \frac{1}{a} \nabla ({\bf v} n_{tr}) = \phi
\ee
where $a$ is the scale factor. Using the expression 
$n_{tr}={\bar n}_{tr}(1+\delta_{tr})$,
in which ${\bar n}_{tr}$ is the mean tracer density (depending on redshift),
then the relation between $\phi$ and our $\Psi$ function is given by:

\be
|\Psi(t)|=\frac{1}{{\bar n}_{tr}(t)} \left[\phi(t)-(1+\delta_{tr}) 
\frac{d{\bar n}_{tr}}{dt}\right]
\ee

Now, we can attempt to explore the functional form of our $\Psi$ function,
taking into account that the evolution of the galaxy luminosity 
function, $\Phi(L)$, can be parametrized ($k\sim 3$) as:
\be
\Phi(L,z){\rm d}L=(1+z)^{k} \Phi(L,0){\rm d}L \;\;.
\ee
Therefore, the corresponding mean density as a function of redshift
becomes:
\be
{\bar n}_{tr}(z)=(1+z)^{k}\int_{L_{min}}^{\infty} \Phi(L) {\rm d}L=
{\bar n}_{tr}(0)(1+z)^{k} \;\;.
\ee
On the other hand, the time derivative of the mean density is written:

\be 
\frac{d{\bar n}_{tr}}{dt}=\frac{d{\bar n}_{tr}}{dz}\frac{dz}{dt}=
-H_{0}k{\bar n}_{tr}(0)(1+z)^{k-1}(1+z)E(z)=-kH_{0}{\bar n}_{tr}(z)E(z) \;\;.
\ee
Therefore we have that
\be\label{eq:ap1}
|\Psi(z)| = \left[ \frac{\phi(z)} {{\bar n}_{tr}(z)}+
kH_{0}(1+\delta_{tr}(z))E(z) \right] \;\;.
\ee
In order to proceed, we make 
the following assumption (similar to that of Simon 2005), 
namely a Taylor expansion in 
${\bar n}_{tr}(z)$ up to the second order:
\be
\phi=A+B{\bar n}_{tr}(z)+C {\bar n}_{tr}^{2}(z)+... 
\ee
with $A, B$ and $C$ constants. Doing so, eq.\ref{eq:ap1} becomes:
\be
|\Psi(z)|\simeq \frac{A}{{\bar n}_{tr}(0)(1+z)^{k}}+B+
C{\bar n}_{tr}(0)(1+z)^{k}+kH_{0}E(z)+kH_{0}E(z)\delta_{tr}(z)
\ee
or 
\be
|\Psi(z)|\simeq (1+z)^{k} \left[ \frac{A}
{ {\bar n}_{tr}(0)(1+z)^{2k}}+
\frac{B}{(1+z)^{k}}
+C{\bar n}_{tr}(0)
+\frac{kH_{0}E(z)}{(1+z)^{k}}+
\frac{kH_{0}E(z)F(z)}{(1+z)^{k+1}} \right] 
\ee
where in the framework of the $\Lambda$ cosmology, we have used that 
\be
\delta_{tr}\propto (1+z)^{-1}
F\left[\frac{1}{3},1,\frac{11}{6},-\frac{\Omega_{\Lambda}}
{\Omega_{\rm m} (1+z)^{3}} \right] 
\ee
(see Silveira \& Waga 1994) with $F(z)$ being the hypergeometric function
which is an increasing function of redshift. Therefore, for $k\sim 3$ and 
at the high redshift limit, we have:
\be\label{eq:ap2}
|\Psi(z)|\longrightarrow (1+z)^{3} \left[C{\bar n}_{tr}(0)+ 
\frac{kH_{0}E(z)}{(1+z)^{3}} \right]\;\;,
\ee
while for the Einstein de Sitter universe:
\be\label{eq:ap3}
|\Psi(z)|\longrightarrow (1+z)^{3} \left[C{\bar n}_{tr}(0)+ 
\frac{kH_{0}}{(1+z)^{3/2}} \right]\;\;.
\ee
It is evident that the second term of the right hand-side of eqs. \ref{eq:ap2} 
and \ref{eq:ap3}
is small at large redshifts, which implies that indeed, as we assumed in our 
model of $\Psi$ (eq. 19 of main paper), $\Psi(z) \propto (1+z)^{\nu}$.
We also show analytically that we should expect $\nu \sim 3$, and indeed we 
find, fitting the simulation data, a similar value, ie., $\nu \sim 2.5$. 

\begin{table}
\caption[]{Fitted {\em Case 2}
bias evolution model parameters and their
$1\sigma$ uncertainties, for different halo-mass ranges.
For the $S_{4}$ and $S_{5}$ samples, the effects of 
halo interactions is negligible and thus 
the {\em Case 1} model is valid.}

\tabcolsep 2pt
\begin{tabular}{cccccccc} 
\hline
Sample & $M/h^{-1}M_{\odot}$& ${\cal C}_{1}$& ${\cal C}_{2}$ & $\chi^{2}/dof$ &$b_{0}$&$A$&$\nu$ \\ \hline \hline 
$S_{1}$& $7.7\times 10^{11}-1.0\times 10^{13}$  &2.12$\pm 0.09$ & -0.24$\pm 0.01$& 1.15&$0.91\pm 0.01$
&$2\pm 0.08 \times 10^{-3}$&$2.56\pm 0.11$\\
$S_{2}$& $1.0\times 10^{13}-3.0\times 10^{13}$  &3.98$\pm 0.10$ & -0.43$\pm 0.01$& 1.30&$0.98\pm 0.02$
&$5\pm 0.20 \times 10^{-3}$&$2.62\pm 0.14$\\ 
$S_{3}$& $3.0\times 10^{13}-6.0\times 10^{13}$  &5.48$\pm 0.85$ & -0.57$\pm 0.09$& 0.50&$1.16\pm 0.04$
&$6\pm 0.30 \times 10^{-3}$&$2.54\pm 0.15$\\ 
$S_{4}$& $6.0\times 10^{13}-1.0\times 10^{14}$  &6.78$\pm 2.80$ & -0.68$\pm 0.30$& 0.20&$1.37\pm 0.10$& &\\ 
$S_{5}$& $1.0\times 10^{14}-4.0\times 10^{14}$  &9.09$\pm 3.00$ & -0.89$\pm 0.50$& 0.20&$1.67\pm 0.10$& &\\ \hline

\end{tabular}
\end{table}

\end{document}